\documentclass[9pt,twocolumn,twoside]{osajnl}

\journal{ol} 

\setboolean{shortarticle}{true} 

\title{Tunable Fluidic Lenses with High Dioptric Power\\ for Impaired Vision}

\author[1,2*]{Graciana Puentes}

\author[1,3]{Fernando Minotti}

\affil[1]{Universidad de Buenos Aires. Facultad de Ciencias Exactas y
Naturales. Departamento de F\'{\i}sica. Buenos Aires, Argentina.}
\affil[2]{CONICET-Universidad de Buenos Aires. Instituto de F\'{\i}sica de Buenos Aires (IFIBA). Buenos Aires, Argentina.}
\affil[3]{CONICET-Universidad de Buenos Aires. Instituto de F\'{\i}sica del Plasma (INFIP). Buenos Aires, Argentina.}

\affil[*]{Corresponding author: gracianapuentes@gmail.com}

\dates{Compiled \today}

\ociscodes{(010.1080) adaptive optics; (220.3620) lens desing}

\doi{\url{http://dx.doi.org/10.1364/ao.XX.XXXXXX}}

\begin{abstract}
We report experimental and theoretical results on the production of macroscopic fluidic lenses with high dioptric power, tunable focal distance and aperture shape, for applications in adaptive eyewear for the sub-normal vision segment.
The lense is 17 mm wide and is made of an elastic PDMS polymer, which can adaptively restore accommodation distance within several centimeters according to the fluidic volume mechanically pumped in. 
Moreover, the lens can provide for magnification in the range of +25 Diopter to +100 Diopter with optical aberrations on the order of the wave-length, and overall lens weight of less than 2 $g$. We argue that these features make the proposed lenses appropriate for the impaired vision segment. 
\end{abstract}

\setboolean{displaycopyright}{true}

\begin{document}

\maketitle
\thispagestyle{fancy}

\ifthenelse{\boolean{shortarticle}}{\ifthenelse{\boolean{singlecolumn}}{\abscontentformatted}{\abscontent}}{}

\section*{Introduction}

National estimates reveal that the vision correction market in the U.S. nowadays is enormous. About seven out of ten Americans 18 years of age and older wear some type of corrective lens, a figure that jumps to $90\%$ and higher among those 50 and older. Corneal refractive errors are typically corrected by permanent means such as glasses, contact lenses or surgery. Standard eyewear corrects for refractive errors, by shifting the focal plane in the case of myopia and hyperopia, or by introducing astigmatic correction in the case of coma, astigmatism and other higher-order aberrations \cite{Gross, Keating, Tasman}. Moreover, in many situations visual impairments are combined, for instance most myopia and hyperopia patients also suffer from presbyopia requiring multi-focal lenses. However, permanent multi-focal lenses limit the accommodation range of the eye and therefore limit the field-of-view to small focal regions suitable to observe object at different distances, thus resulting in a considerable vision loss \cite{Johnson}. Currently, visual impaired patients, such as the result of glaucoma or other visual trauma, have to rely on organic glasses of large dioptric power. Large dioptric power with standard lenses can only be achieved with thick and heavy components, which introduce a large number of optical aberrations, make the glasses very bulky and unpractical. Moreover, the employed standard lenses are static and can only provide for a fixed focal distance. 

To our knowledge, all previous prototypes for adaptive eyewear \cite{WO2006,Mastrangelo} focus on different segments of the vision market, with moderate augmentation requirements. In particular, in Ref. \cite{WO2006} the author propose a fluidic lens prototype with optical power in the range (+/-6 D to +/- 20D). On the other hand in Ref. \cite{Mastrangelo}, the authors propose a fluidic lens prototype with optical power in the range (+/-3D to +/-4 D).  In this letter, we present the first macroscopic fluidic lens eyewear prototype with high dioptric power (+25D to +100D range) with optical aberrations in the order of the wave-length oriented to the impaired vision market, which can adaptively restore accommodation distance within several centimeters, thus enabling access to the entire field-of-view. The lens is made of an elastic polymer which can adaptively modify its optical power according to the fluid volumen mechanically pumped in. Such liquid lens exhibits a large dynamic range, and its focussing properties are polarization independent  \cite{Puentes}. Additionally, we demonstrate that by tuning the lens aperture it is possible to address different optical aberrations, thus providing an additional degree of freedom for the lens design. Our design is not only attractive for adaptive eyewear, but also for cellular phone, camera, and other machine vision applications where large augmentation can be required. \\

Adaptive fluidic lenses can be classified in two main categories, according to their operation mechanism. The first class is the electro-wetting lens whose focal length can be tuned continuously by applying a controlled external voltage \cite{Vallet, Krupenkin, Kuiper}. While the advantage of electro-wetting lenses is that they can provide for large focusing power with no mechanical moving parts, the disadvantage is a relatively high driving voltage required, in addition to limited stability and liquid evaporation. The second type is the mechanical lens, whose focal length is controlled by pumping liquid in and out of the lens chamber, which changes the curvature of the lens profile \cite{Knollman, Sugiura, Zhang, Jeong, Chen, Chronis, Moran, Ren}. The adaptive lenses reported here belong to the second category. 

\section*{Theoretical Model}

To study the nonlinear, large deformation of a thin elastic membrane we
employ the equations derived by Berger \cite{berger} for thin isotropic
elastic plates: 
\begin{equation}
\label{b1}
\nabla ^{4}w-\alpha ^{2}\nabla ^{2}w =\frac{q}{D},  
\end{equation}
\begin{equation}
\label{b2}
\frac{\partial u}{\partial x}+\frac{\partial v}{\partial y}+\frac{1}{2}
\left( \frac{\partial w}{\partial x}\right) ^{2}+\frac{1}{2}\left( \frac{
\partial w}{\partial y}\right) ^{2} =\frac{\alpha ^{2}h^{2}}{12}.
\end{equation}
In these equations $w\left( x,y\right) $ is the local $z$-displacement of
the membrane, with non-deformed state corresponding to the $z=0$
plane, $u\left( x,y\right) $ and $v\left( x,y\right) $ are the local $x$ and 
$y$ displacements, $D$ the membrane bending rigidity, and $h$ its thickness.
The magnitude $q$ corresponds to the applied $z$-load, and $\alpha $ is a
constant to be determined from the same equations by imposing appropriate
boundary conditions. Using the method of Mazumdar \cite{mazumdar1} it is possible to obtain an ordinary differential equation for $w$ if the expression of the contour lines of equal deflection, $\psi \left(
x,y\right) $, is known. In that case one integrates Eq. (\ref{b1})
over the area $S_{\psi }$ enclosed by the contour line corresponding to a
given value of $\psi $. Since by definition $w=w\left( \psi \right) $, one
obtains from Eq. (\ref{b1}), for the case of uniform load $q$,

\begin{eqnarray}
\frac{q}{D}S_{\psi } &=&\frac{d^{3}w}{d\psi ^{3}}\oint\limits_{\psi }\left(
\psi _{x}^{2}+\psi _{y}^{2}\right) ^{2}\frac{\left\vert dx\right\vert }{
\left\vert \psi _{y}\right\vert }+\frac{d^{2}w}{d\psi ^{2}}
\oint\limits_{\psi }\left[ \left( \psi _{x}^{2}+\psi _{y}^{2}\right) \nabla
^{2}\psi \right.   \nonumber \\
&&\left. +2\left( \psi _{x}^{2}\psi _{xx}+2\psi _{x}\psi _{y}\psi _{xy}+\psi
_{y}^{2}\psi _{yy}\right) \right] \frac{\left\vert dx\right\vert }{
\left\vert \psi _{y}\right\vert }  \nonumber \\
&&+\frac{dw}{d\psi }\oint\limits_{\psi }\left( \psi _{x}\nabla ^{2}\psi
_{x}+\psi _{y}\nabla ^{2}\psi _{y}\right) \frac{\left\vert dx\right\vert }{
\left\vert \psi _{y}\right\vert }  \nonumber \\
&&-\alpha ^{2}\frac{dw}{d\psi }\oint\limits_{\psi }\left( \psi _{x}^{2}+\psi
_{y}^{2}\right) \frac{\left\vert dx\right\vert }{\left\vert \psi
_{y}\right\vert },  \label{b1psi}
\end{eqnarray}
where $x$ and $y$ subindexes represent partial derivatives with respect to
the corresponding variable.

It is convenient to assign the value $\psi =0$ to the boundary of the
membrane, so that a boundary condition is $w\left( 0\right) =0$. For the
case of a clamped membrane, another condition is $dw/d\psi =0$ at $\psi =0$.
Since the differential equation is of third order an additional condition is
required, which results generally from considerations of regularity of the
solution.

On the other hand, Eq. (\ref{b2}) can be written, for the case of no
pre-stretching ($u=v=0$ at the clamped edge of the membrane), as 
\begin{equation}
\frac{\alpha ^{2}h^{2}}{6}S_{0}=\int_{0}^{1}d\psi \left( \frac{dw}{d\psi }
\right) ^{2}\oint\limits_{\psi }\left( \psi _{x}^{2}+\psi _{y}^{2}\right) 
\frac{\left\vert dx\right\vert }{\left\vert \psi _{y}\right\vert }.
\label{alpha}
\end{equation}

The method considered requires to know the expression of $
\psi \left( x,y\right) $. In the case of an elliptical membrane of 
$x$, $y$ semi-axes $a$ and $b$, such an expression can be obtained considering that for uniform load equal deflection contour lines are expected to follow the
shape of the membrane boundary at $\psi =0$ and thus correspond to the
ellipses: 
\begin{equation}
\psi \left( x,y\right) =1-x^{2}/a^{2}-y^{2}/b^{2}.  \label{psi_ellipse}
\end{equation}

Use of (\ref{psi_ellipse}) in Eq. (\ref{b1psi}) results in
\begin{equation}
\left( \psi -1\right) \frac{d^{3}w}{d\psi ^{3}}+2\frac{d^{2}w}{d\psi ^{2}}
+\gamma ^{2}\frac{dw}{d\psi }=Q,  \label{eq_w}
\end{equation}
where
\begin{eqnarray}
\gamma ^{2} &=&\alpha ^{2}\frac{a^{2}b^{2}\left( a^{2}+b^{2}\right) }{
3a^{4}+2a^{2}b^{2}+3b^{4}},  \label{gamma2} \\
Q &=&\frac{a^{4}b^{4}}{3a^{4}+2a^{2}b^{2}+3b^{4}}\frac{q}{2D}.  \label{big_Q}
\end{eqnarray}

The solution of Eq. (\ref{eq_w}) with boundary conditions $w=dw/d\psi =0$ at 
$\psi =0$ is easily obtained changing to the variable \cite{mazumdar2} 
\begin{equation}
\varsigma ^{2}=1-\psi ,  \label{zeta_2}
\end{equation}
resulting in 
\begin{equation}
w\left( \varsigma \right) =\frac{Q}{\gamma ^{3}I_{1}\left( 2\gamma \right) }
\left[ \gamma \left( 1-\varsigma ^{2}\right) I_{1}\left( 2\gamma \right)
+I_{0}\left( 2\gamma \varsigma \right) -I_{0}\left( 2\gamma \right) \right] ,
\label{w_zeta}
\end{equation}
where $I_{n}$ is the order $n$ modified Bessel function of the first kind.
In the determination of (\ref{w_zeta}) the condition of finite values of $w$
at $\varsigma =0$ was also employed. The value of $\gamma $ is determined,
from the Eqs. (\ref{alpha}), (\ref{psi_ellipse}) and (\ref{gamma2}), as the
solution to the relation 
\begin{equation}
\gamma ^{2}=\frac{6}{h^{2}}\frac{\left( a^{2}+b^{2}\right) ^{2}}{%
3a^{4}+2a^{2}b^{2}+3b^{4}}\int_{0}^{1}\left( \frac{dw}{d\varsigma }\right)
^{2}\varsigma d\varsigma .  \label{gamma2_ellipse}
\end{equation}

\section*{Fluidic Lens Prototype and Experimental Results}

\begin{figure}[h!]
\centering
\includegraphics[width=0.9\linewidth]{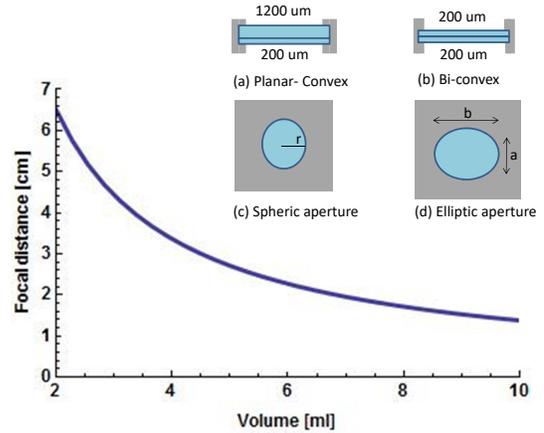}
\caption{Inset: Scheme of fluidic lenses. By tuning the width of the elastic membrane it is possible to design (a) planar-convex and (b) bi-convex lenses. By changing the shape of the aperture between (c) spheric or (d) elliptical,
 it is possible to address different visual aberrations. Blue curve shows a simulation of  the expected focal distances vs. fluidic volumen for a spheric aperture ($r=17$ mm), based on theoretical model. The obtained focal distance can be tuned across 1cm to 5cm. 
 This focal distance range  corresponds to an optical power range of +25D to +100D, thus confirming the high dioptric power and dynamic range of the fludic lenses.  }
\label{fig:false-color}
\end{figure}

The lens consists of two layers made of an elastic membrane of the polydimethylsiloxane (PDMS) type, which was prepared at Laboratory of Polymers (University of Buenos Aires). The fabrication is straightforward since it is possible to use a master mold able to repeat the procedure with precision of 0.1 um to 10 nm fidelity \cite{Polson}. 
 By choosing a different thickness for the elastic membranes we can manufacture plano-convex (Fig. 1 (a)) or bi-convex lenses (Fig. 1 (b)). For the plano-convex case one of the membranes is made significantly thicker than the other, in order to remain uncurved under pressure. Typical thickness for the thick membrane is 1200 $\mu m$, and for the thin membrane 200 $\mu m$. The two elastic films are held together by an aluminum frame, sealed with the elastic membrane. A fluid  of refractive index matched to the polymer ($n=1.47$) such as glycerol or distilled water is injected between the elastic layers. By increasing or decreasing the fluid volume injected, it is possible to tune the focal distance across several centimeters, and therefore adjust the optical power of the lens. 
Further, we tune one additional degree of freedom, given by the shape of the aperture. By modifying the aperture shape from spherical (Fig. 1 (c)) to elliptical (Fig. 1 (d)) we can introduce different optical corrections. Typical size for the spherical lens is given by a diameter of 17mm, the elliptic lens has a large axis $a=17$ mm, and a small axis $b=15,13$ and 12 mm. 
Using the theoretical model for the deformation in the elastic membrane as a function of the liquid volume, we simulated the expected focal distances of the fluidic lenses as a function of volume. Experimental results and theoretical model are displayed in Fig. 2, the agreement between theory and experiment is apparent. The focal distance range is 1cm-5cm. Using the standard conversion (1000mm/D=f), we obtain and optical power range of +25D to +100D, clearly demonstrating that our prototype operates in at a different dynamic range, not explored in previous designs.

\begin{figure}[t!]
\centering
\includegraphics[width=0.8\linewidth]{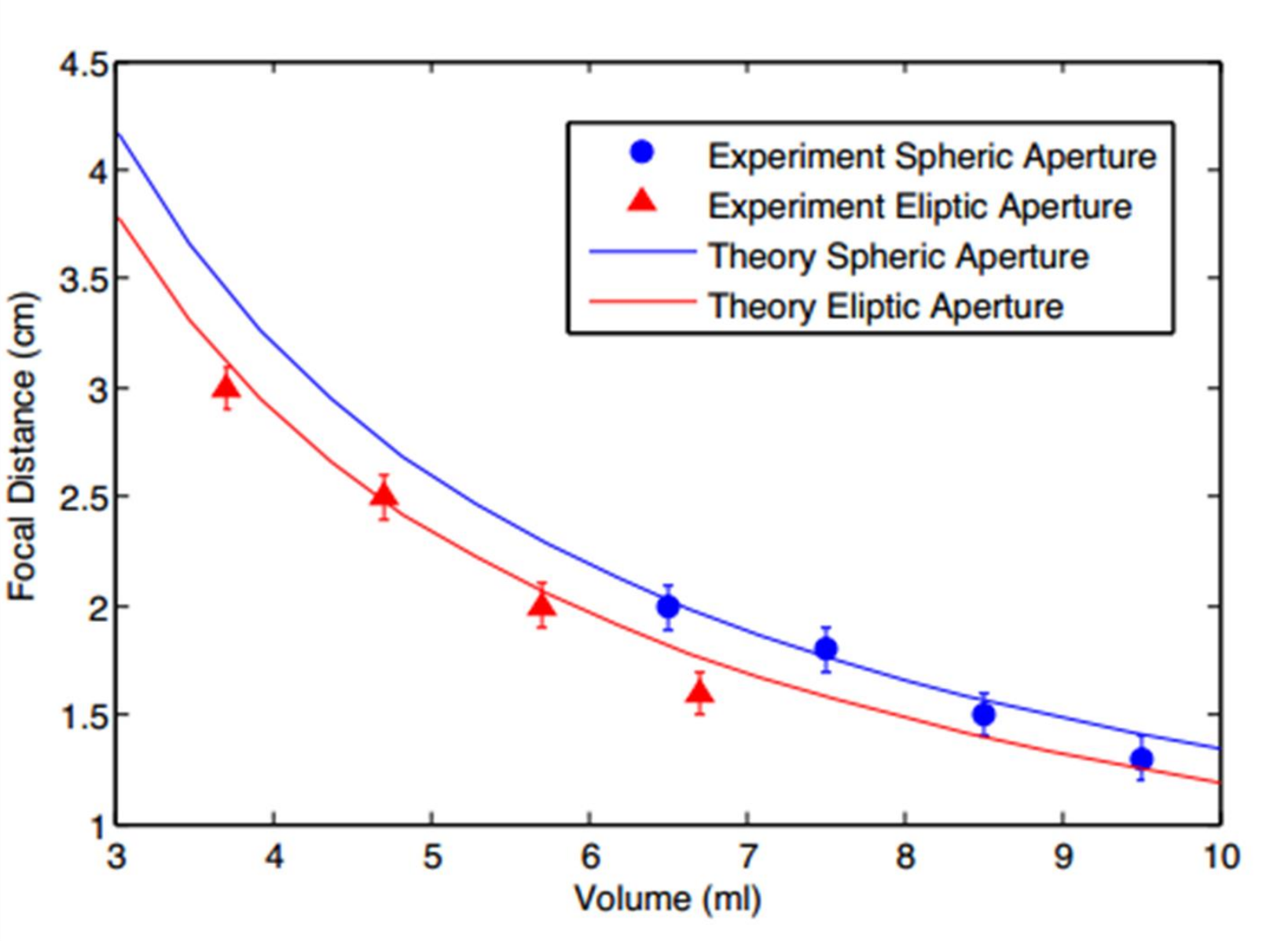}
\caption{Measured focal distance vs. fluidic volume for a lens with elliptic aperture (red triangles) and spheric aperture (blue circles). Blue curve: Theoretical prediction for a spheric lens with 17mm  radius. Red curve: Theoretical prediction for an elliptic lens with mayor and minor axes 17 
mm and 15 mm, respectively. }
\label{fig:false-color}
\end{figure}

Tunable optical power is a very desirable property for any user-oriented device. In particular, for eyewear lenses, cameras or other machine vision applications. One main applications is for instance in correcting for presbyopia. 
Typical multi-focal crystal lenses, are static and can only contain a limited number of focal distances, making the field-of-view of the lens itself very narrow.
 The tunable lens we demonstrate here can provide for a continuum of focal distances expanding across several centimeters without compromising the field-of-view, making them highly suitable for presbyopia applications where typical mutli-focal lenses have a focal distance
  variation of 32mm, among others. To characterize the focusing properties of the lens we use an expanded He-Ne laser beam as a probing light source, and measure the position of focussing spot.  We measured the focuss dynamic range of the fabricated prototype by increasing the fluid volume from an initial arbitrary value $V_0$. In our demonstration we inreased the fluid volume mechanically in steps of 1 ml, however this fluid volume can be easily controlled via voltage, motors, or electronic means. The measured dynamic range is displayed in Fig. 2, and is larger than 3 cm, which corresponds to a dynamic range of 75 Diopters. We measured the optical power of the lenses, for the case of spherical aperture and elliptic aperture. By changing the fluid volume the optical power and focal distance tuned in the  expected range, as predicted by the model.

\section{Lens Aberration Analysis}

In order to determine the aberration effects of a single membrane we
consider a plane wave front parallel to the $z=0$ plane incident on the
membrane from its internal side. The corresponding rays, parallel to the $z$
-axis, with direction $\mathbf{e}_{z}$, are then refracted according to
Snell law when they cross the membrane surface at $z=w\left( x,y\right) $.
The external normal unit vector of the membrane surface is given by (in
cartesian components) 
\begin{equation}
\mathbf{n}=\frac{\left( -w_{x},-w_{y,}1\right) }{\sqrt{1+w_{x}^{2}+w_{y}^{2}}
}.
\end{equation}
Use of Snell law thus results in the refracted ray having the direction given by the unit vector 
\begin{equation}
\mathbf{k}_{r}=\mathbf{n}\cos \theta _{r}+\mathbf{t}\sin \theta _{r},
\end{equation}
where 
\begin{equation}
\sin \theta _{r}=n_{f}\frac{\sqrt{w_{x}^{2}+w_{y}^{2}}}{\sqrt{
1+w_{x}^{2}+w_{y}^{2}}},
\end{equation}
and
\begin{equation}
\mathbf{t}=\frac{\left( w_{x},w_{y,}w_{x}^{2}+w_{y}^{2}\right) }{\sqrt{
\left( 1+w_{x}^{2}+w_{y}^{2}\right) \left( w_{x}^{2}+w_{y}^{2}\right) }}.
\end{equation}

In this way, a generic ray refracted at the point $\mathbf{X}_{0}=\left(
x,y,w\left( x,y\right) \right) $ on the membrane surface, intersects a plane 
$z=d$ at the point $\mathbf{X}_{d}=\mathbf{X}_{0}+\mathbf{k}_{r}L$, with 
\begin{equation}
L=\frac{d-w\left( x,y\right) }{k_{rz}},
\end{equation}
and so the phase distribution on the plane $z=d$, corresponding to a front
of phase $\phi _{0}$ at $z=0$ is ($\lambda $ is the wavelength in air) 
\begin{equation}
\phi _{d}=\phi _{0}+\frac{2\pi }{\lambda }\left[ n_{f}w\left( x,y\right) +
\frac{d-w\left( x,y\right) }{k_{rz}}\right] .  \label{phase}
\end{equation}

This is displayed in Fig. 3.
In order to quantify the aberrations in terms of Zernike coefficients we consider the difference between the phase (\ref{phase}) and the phase at the same plane $z=d$ corresponding to the same front, but refracted by an ideal lens of focal distance equal to the mean focal distance of the central region of the fluidic lens, $f_{0}$ (see Fig. 4 and Table 1),
\begin{equation}
\phi _{ideal}=\phi _{0}+\frac{2\pi }{\lambda }\left[ n_{f}w\left( 0,0\right)
+f_{0}+D\left( x,y\right) \right] ,
\end{equation}
where $D\left( x,y\right) $ is the distance from the position of $f_{0}$ to
the point $\left( x_{d},y_{d}\right) $ determined by $\mathbf{X}_{d}=\mathbf{
X}_{0}+\mathbf{k}_{r}L$. In this way the effects due to analyzing a curved wavefront at a plane surface are subtracted, and only true aberration effects are left. The resulting coefficients become independent of the position $d$ of the analyzing plane for $d\gg f_{0}$.

\begin{figure}[t!]
\centering
\includegraphics[width=1\linewidth]{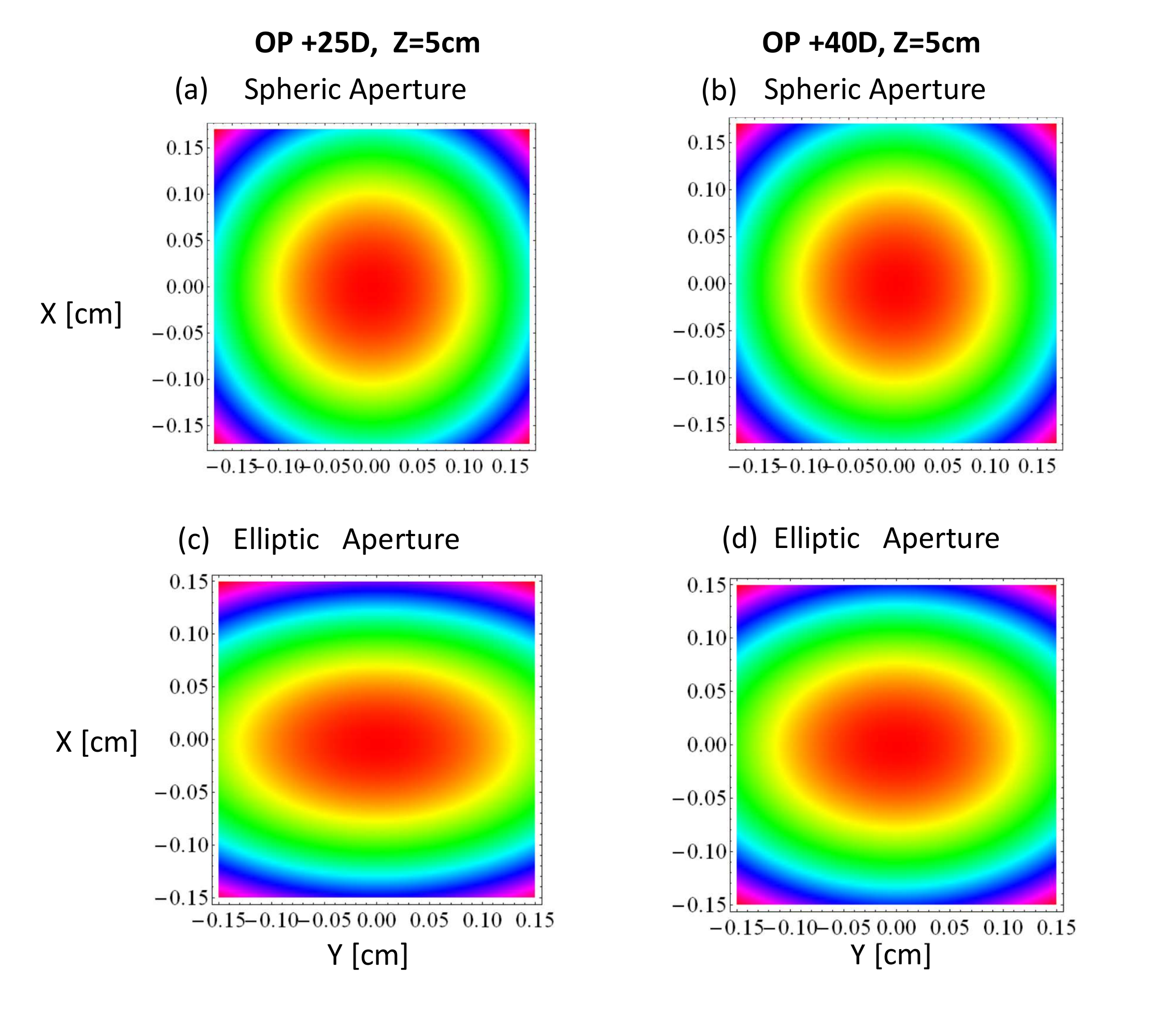}
\caption{Simulated phase front for paraxial rays and propagation distance $z=5$ cm. The diameter of the aperture is 3mm, which is the typical aperture diameter for a Shack-Hartmann Wave-Front Sensor \cite{Mastrangelo}.  Left column: Fluidic Lens with +25D optical power, right column: Fluidic lens with +40D
 optical power. Upper row: Spherical aperture with $r=17$ mm, lower row: Elliptic aperture with minor axis $a=15$ mm, and major axis $b=17$ mm. }
\label{fig:false-color}
\end{figure}

\begin{table}[b!]
\centering
\caption{Simulated optical aberrations in the visible in terms of Zernike polynomials. Table displays the coefficients of Zernike polynomials up to 8-$th$ order. We consider fluidic lenses with optical power +25D and +40D, respectively. We  also consider either spheric or elliptic aperture. For spheric lenses with $r=17$ mm all aberrations are within the order of the wave-length. For elliptic lenses with minor axis $a=15$mm and major axis $b=17$ mm aberrations can be tuned to arbitrary values, and could be used to compensate for ophthalmic refractive errors. Values reported with accuracy up to 1 nm.}
\label{my-label}
\begin{tabular}{lllll}
\hline
 Aberration [$\mu$m] & OP=+25 D $r=17$ mm & OP=+40D, $r=17$mm  & OP=+25D, $a=15$mm  &  OP=+40D, $a=15$mm\\
 \hline
 \hline
 Astigmatism 45 & 0.000 & 0.000 & 0.000 & 0.000 \\
 Astigmatism 90 & 0.000 & 0.000 & -3.240 & -5.236
\\
 Tilt X& 0.000 & 0.000 & 0.000 &0.000\\
 Tilt Y&  0.000 & 0.000 & 0.000 & 0.000 \\
 Coma X& 0.000 & 0.000 & 0.000 & 0.000 \\
 Coma Y& 0.000 &  0.000 & 0.000 & 0.000 \\
 Spherical & -0.009 & -0.037 & -0.007 & -0.029\\
 Field curvature & -0.026 & -0.109 & -0.440 & -0.740\\
 \hline
\end{tabular}
\end{table}

\section*{Conclusions}

We have reported on the design and construction of macroscopic fluidic lenses with high dioptric power, and tunable focal distance for the impaired vision segment. A numerical analisis of the lens aberrations demonstrate that optical aberrations are in the order of the wavelength.
 In addition the lenses are extremely light and affordable, we believe all of this indicates the proposed prototype is highly suitable for the sub-normal vision segment.

\section*{Acknowledgements}
The authors are grateful to the Solar Energy Department (TANDAR-CNEA)  and to the Laboratory of Polymers (FCEN-UBA) for assistance in PDMS membrane preparation. GP and FM gratefully acknowledge financial support from UBACYT 2016 Mod I 20020150100096BA, 
PIP GI 11220120100453, PICT2014-1543, PICT2015-0710 Startup, UBACyT PDE 2015, UBACyT PDE 2017.

\end{document}